# The evidence contained in the *p*-value is context-dependent

Florian Hartig[1,*] and Frédéric Barraquand[2]

Theoretical Ecology Lab, University of Regensburg, Regensburg, Germany, @florianhartig

[2] Institute of Mathematics of Bordeaux, CNRS and University of Bordeaux, Talence, France

@FredBarraquand

* Corresponding Author, florian.hartig@ur.de

In a recent opinion article, Muff et al. [1] recapitulate well-known objections to the Neyman-Pearson Null-Hypothesis Significance Testing (NHST) framework and call for reforming our practices in statistical reporting. We agree with them on several important points: the significance threshold $p<0.05$ is only a convention, chosen as a compromise between type I and II error rates [cf. 2]; transforming the *p*-value into a dichotomous statement leads to a loss of information [cf. 3]; and *p*-values should be interpreted together with other statistical indicators, in particular effect sizes and their uncertainty [4]. In our view, a lot of progress in reporting results can already be achieved by keeping these three points in mind.

We were surprised and worried, however, by Muff et al.'s suggestion to interpret the *p*-value as a "gradual notion of evidence". Muff et al. recommend, for example, that a *p*-value > 0.1 should be reported as "little or no evidence" and a *p*-value of 0.001 as "strong evidence" in favor of the alternative hypothesis H1.

We have multiple reasons for concern with this proposal. First, for all its faults, `statistical significance' ($p<\alpha$) is now a term with a well-defined meaning for researchers with appropriate statistical training. For example, it is usually stressed to researchers in their initial statistical training that neither the *p*-value nor the term significance can be translated into the probability of an effect. The term "strong evidence", on the other hand, has no comparably sharp definition [5], and researchers have not been trained to avoid possible misinterpretations. If results were reported as recommended by Muff et al., we anticipate that most scientific readers would interpret the statement "both studies provide strong evidence for an effect" as saying that, based on each study alone, it is equally likely that an effect exists. Unfortunately, however, this intuitive interpretation of the word "evidence" does not map to the information contained in the *p*-value.

To understand why, consider that for underpowered experiments, p-distributions are approximately flat under either H0 and H1, whereas in high-powered experiments, we are much more likely to observe low p-values under H1 (Fig 1A, see also [6]). As a consequence, the probability that a significant *p*-value is a false positive (= false discovery rate) or below any other threshold depends on the power of the experiment, a point frequently noted in discussions of the replication crisis and the shortcomings of the NHST framework. For these reasons and more, [7] warn that "Researchers should recognize that a *p*-value without context or other evidence provides limited information" and "By itself, a *p*-value does not provide a good measure of evidence regarding a model or hypothesis".

Muff et al. [8] respond to similar points from Lakens [6] by saying that evidence labels "should not be understood as hard thresholds" and that "most applied scientists would agree that it matters a lot whether *P* = 0.001 or *P* = 0.999.". While this may be true, it fails to address the criticism that the probability of obtaining a low *p*-value under H1, and thus the evidence contained in the p-value, is inherently power-dependent.

When looking for a statistical indicator that maps better and context-free to the intuitive linguistic interpretation of "evidence", one would be hard-pressed to not consider the Bayes Factor (BF), defined as Pr(D|H1) / Pr(D|H0) [9]. The BF expresses the relative support for the two hypotheses from the data D, and

assuming prior odds of 50:50, we can convert the BF into the posterior probability of an effect Pr(H1|D). Under certain restrictive assumptions, it is possible to establish a correspondence between the *p*-value and the BF [10], but our simulations illustrate the well-known fact that the relationship between evidence expressed as Pr(H1|D) and the *p*-value is sample-size dependent (Fig 1B), reinforcing the intuition that it is not possible to interpret the evidence contained in a *p*-value without considering sample size.

All this is not saying that the *p*-value does not correlate with evidence. All other things equal, a lower *p*-value means more evidence against the null. Importantly, however, the reverse is also true: the higher the *p*-value, the more evidence in favor of the null. In our example, most observations of *p*>0.1 would translate into Pr(H0|D) = 1-Pr(H1|D) > 50% (Fig 1B). Interpreting this as "no evidence against the null", as suggested by Muff et al., also seems misleading.

In conclusion, Muff et al. propose to attach verbal labels to the well-known star notation (ns, ., *, **, ***) that is already commonly used in statistical reporting. It would be convenient for scientists if such a direct translation of *p*-values into verbal statements about the evidence for an effect existed, as those statements would simplify the notoriously difficult interpretation of the *p*-value considerably. Unfortunately, however, the evidence contained in a *p*-value is context dependent, and suggesting otherwise, if only verbally, only invites further misinterpretations of this famously misunderstood quantity [11].

We therefore advise against using the suggested interpretation of p-values as evidence in statistical reporting. If researchers desire to avoid the significance cut-off, they may just report the *p*-value numerically. If they want to express the evidence in favor of H1, we recommend calculating the Bayes Factor, which has limitations as well (in particular its dependence on the parameter priors), but those limitations are well understood and can be consistently interpreted.

https://github.com/florianhartig/Muff2021Reply

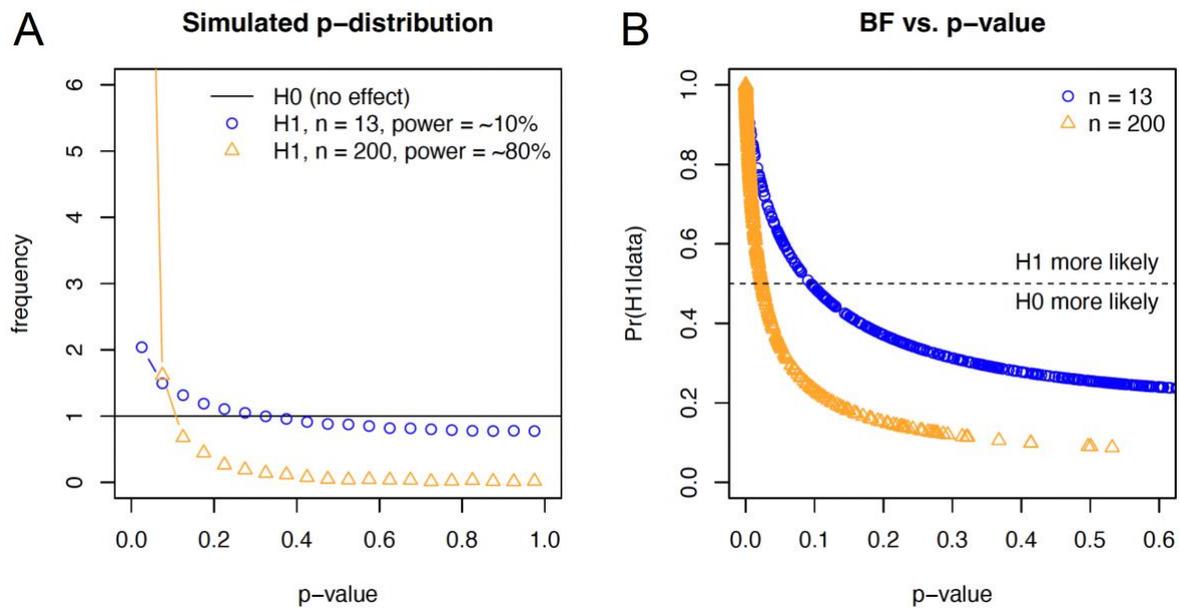

Fig. 1. A) frequency distribution of *p*-values for a 1-sample t-test under the null (black line, we show the theoretical expectation of a uniform distribution), for an underpowered study (blue, simulated, power ~ 10%) and for a well-powered study (orange, simulated, power ~ 80%). B) Posterior probability in favor of H1 calculated using the BayesFactor package with default parameter priors and even prior odds for H0/H1, for the same scenario. For code to reproduce the figure, see https://github.com/florianhartig/Muff2021Reply.